\documentclass[10pt,a4paper]{article}

\usepackage{xcolor}
\usepackage{changebar}
\usepackage{graphicx}

\title{Tapered Yagi-Uda Nanoantennas for Broadband Unidirectional Emission}

\author{Isabelle Staude$^{1,*}$, Ivan S. Maksymov$^{1}$, Manuel Decker$^1$, \\ Andrey E. Miroshnichenko$^1$, Dragomir N. Neshev$^1$, \\ Chennupati Jagadish$^2$, and Yuri S. Kivshar$^1$}

\begin{document}
\maketitle

\noindent $^1$Nonlinear Physics Centre and Centre for Ultrahigh Bandwidth Devices for Optical Systems
(CUDOS), Research School of Physics and Engineering, \\ The Australian National University, Canberra ACT 0200, Australia\\\\
$^2$Department of Electronic Materials Engineering, Research School of Physics and Engineering, The Australian National University, Canberra ACT 0200, Australia\\

\noindent $^*$ips124@physics.anu.edu.anu\\ 

\begin{abstract}
We demonstrate experimentally the operation of tapered Yagi-Uda nanoantennas for broadband unidirectional emission enhancement. The measured transmittance spectra show that, in comparison to untapered reference structures, the tapered nanoantennas exhibit distinct wide-band spectral resonances. The performed full-vectorial numerical calculations are in good qualitative agreement with the measured spectra, further revealing how the near-field profiles of the tapered nanoantennas are directly reflecting their broadband characteristics.
\end{abstract}


\section{Introduction}
\vspace{-1mm}

\begin{figure}[b]
  \centering
  \includegraphics[width=8cm]{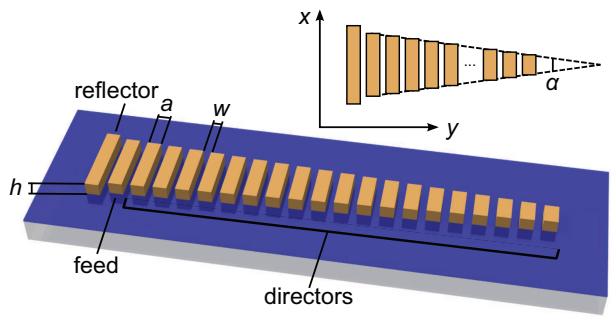}
  \vspace{-2mm}
  \caption{Schematic view of a tapered Yagi-Uda nanoantenna. The inset indicates the orientation of the coordinate system and illustrates the definition of the taper angle $\alpha$.}
  \vspace{-3mm}
  \label{scheme}
\end{figure}

Plasmonic nanoantennas have become a subject of considerable theoretical and experimental interest~\cite{gia11_review, novotny_vanhulst_2011, biagiony_hecht_2012}. Numerous intriguing potential applications of nanoantennas have been considered in areas as diverse as optical and quantum communication, nonlinear optics, sensing, and photovoltaics~\cite{biagiony_hecht_2012}. Arrayed nanoantennas like Yagi-Uda architectures downscaled to nanometer dimensions are particularly suited for these applications because they offer high directivity and strong emission enhancement at the same time~\cite{curto_vanhulst_2010}.

However, a classical design of Yagi-Uda nanoantennas leads to intrinsically narrowband devices, creating the need for new antenna concepts able to provide broadband functionality. Recently, it was shown that a simple design of tapered nanoantenna may lead to strong directionality of emission and broad spectral bandwidth of optical nanoantennas~\cite{mak11_1,mir11}.  Such broadband nanoantennas will open up novel opportunities for high-bandwidth wireless on-chip communication and spectroscopy~\cite{maksymov_kivshar_2012}, multichannel sensing~\cite{maksymov_kivshar_2012c}, and light harvesting devices~\cite{pavlov_vanhulst_2012}. In this paper, we demonstrate the experimental realization of this new design of tapered Yagi-Uda optical nanoantenna, and confirm its directionality over a large spectral bandwidth.

\section{Nanoantenna design}
\vspace{-1mm}
Our nanoantenna design results from applying the concept of tapered plasmonic waveguides to plasmonic Yagi-Uda nanoantennas~\cite{mak11_1}. A schematic view of the considered nanoantenna design is shown in Fig.~\ref{scheme}. The nanoantenna consists of a reflector, a principal feed element, and an array of equally spaced directors. The length of the feed element is adjusted to be resonant at the principal emitted wavelength of the nanoantenna. The reflector element has a length of $L_r=1.25L_f$, and is thus inductively detuned with respect to the feed element. Here $L_f$ is the length of the feed element. The length of the director elements is slowly tapered along the length of the antenna. For a linear taper the length $L_N(\alpha)$ of the $N$-th director is a function of the tapering angle $\alpha$ and is given by the formula $L_N(\alpha)=L_f-2N(w+a)\tan(\alpha/2)$, where $a$ is the width of the elements and $w$ is the distance between neighboring antenna elements. The center-to-center distance of adjacent elements is $w+a=80\rm\,nm$ and the height of the antenna elements is $h=50\rm\,nm$. Compared to classical design principles this means that the spacing is drastically reduced and reaches deep-subwavelength dimensions [$(w+a)\,\approx\,\lambda_0/19$ for a principal emitted wavelength of $\lambda_0=1550\,\rm nm$]. As such our tapered antenna is more compact than e.g., grating-based antennas~\cite{bar11}, which is essential for high-density integration on a photonic chip.

The directionality and the broadband properties of these nanoantennas were theoretically studied in our previous works~\cite{mak11_1, mir11, maksymov_kivshar_2012b}, where the antenna elements were made of silver and suspended in air. It was shown that a pronounced maximum of the directivity and a clear minimum of the beamwidth are simultaneously observed for an optimal taper angle of $\alpha_{opt}=6.6^{\circ}$~\cite{mak11_1}, when the nanoantenna is excited with a point-like dipole emitters placed in the hot spot of the principal feed element. Remarkably, these results are stable against fabrication errors, which makes this nanoantenna design a feasible candidate for real-life applications. Additional simulations assuming a director element length variation defined by $L_N/L_{N+1}=0.99$ furthermore show that this type of nanoantenna offers exceptional broadband functionality. For example, when choosing $L_f=388$\,nm, unidirectional emission enhancement is achieved in a \textit{broad operating band} between $1.32\,\mu\rm m$ and $1.65\,\mu\rm m$ with average front-to-back ratio~\cite{bar11} of $\rm FBR\,\approx\,20$ and radiation efficiency $\eta\,\approx\,20\%$~\cite{mir11}. Moreover, at the principal telecom wavelength of $1.55\,\mu\rm m$ these nanoantennas exhibit theoretical $\rm FBR=85$ accompanied by a radiation efficiency $\eta$ larger than $30\%$.

\begin{figure}[t]
  \centering
  \includegraphics[width=\textwidth]{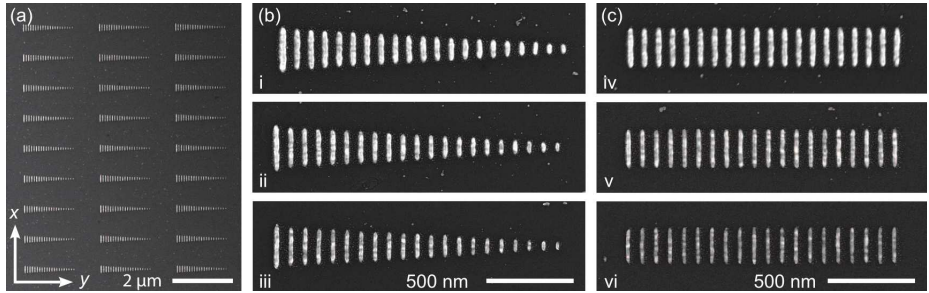}
  \vspace{-2mm}
  \caption{(a) Overview electron micrograph of fabricated nanoantenna arrays. (b) Close-ups of tapered nanoantennas (i-iii) and (c) reference structures (iv-vi) for different exposure doses during electron-beam nanolithography.}
  \vspace{-3mm}
  \label{fabrication}
\end{figure}

\begin{figure}[t]
  \centering
  \includegraphics[width=\textwidth]{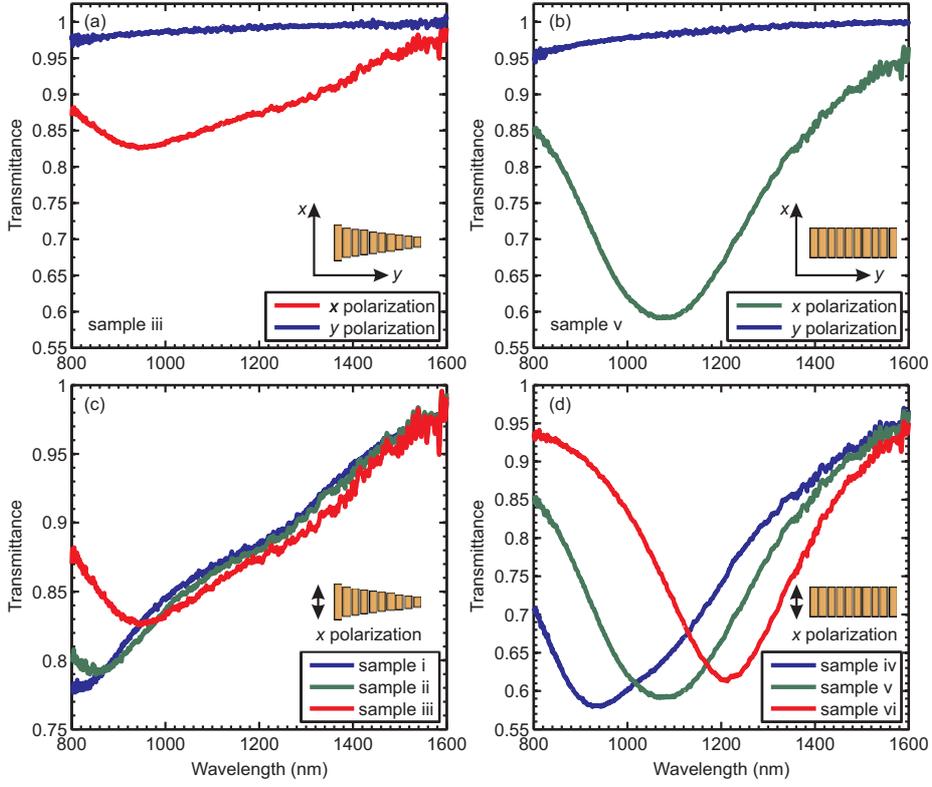}
  \vspace{-2mm}
  \caption{Measured linear-optical transmittance spectra for the incident light polarized in $x$ and $y$ direction (a) of an array of tapered nanoantennas (sample iii) and (b) of an array of reference structures (sample v). Measured linear-optical transmittance spectra for $x$-polarized light of arrays of (c) tapered nanoantennas and (d) reference structures for different widths of the antenna elements.}
  \vspace{-3mm}
  \label{characterisation}
\end{figure}

\section{Experimental results}
\vspace{-1mm}
In experiment we fabricate nanoantennas consisting of $21$ nanorods using elec\-tron-beam nanolithography followed by evaporation of 50\,nm of gold and a lift-off procedure. Gold is used instead of silver to avoid oxidation of the structure under ambient conditions. As a substrate we have used glass covered by 7\,nm of Indium-Tin-Oxide (ITO). In order to allow for assessment of the optical sample quality by means of far-field spectroscopy we have arranged a large number of nominally identical antennas in a two-dimensional array [see a scanning-electron micrograph in Fig.~\ref{fabrication}(a)]. The center-to-center distance between neighboring antennas is $\rm 1\,\mu\rm m$ in $x$-direction and $\rm 2.5\,\mu\rm m$ in $y$-direction. We have chosen the length of the feed element to be 235\,nm and an optimal taper angle of 6.6~degrees in order to tune the antenna's far-field signature to appear within the range of high sensitivity of our characterization setup. The lengths of all other antenna elements are scaled accordingly. Fig.~\ref{fabrication}(b) shows close-ups of typical individual nanoantennas (i-iii) for different nanorod widths corresponding to different exposure doses during electron-beam nanolithography. As a reference we have also fabricated arrays of the same antenna structure but without the reflector and without the taper. These structures allow for an easier theoretical assessment while posing the same experimental challenges, in particular the very small center-to-center distance of adjacent nanorods of only 80\,nm. Close-up electron micrographs of such reference structures (iv-vi) for three different nanorod widths are shown in Fig.~\ref{fabrication}(c).

In order to assess the far-field optical properties of the fabricated structures we collect normal-incidence linear-optical transmittance spectra using a white-light spectroscopy setup and an optical spectrum analyzer. Figure~\ref{characterisation}(a) shows the transmittance through the array of tapered nanoantennas (sample iii) for the two orthogonal linear polarizations of the incident light. A broad resonance with a distinct non-single-Lorentzian shape can clearly be identified for the incident light polarized in $x$-direction. In contrast, the transmittance for incident light polarized in $y$-direction is close to unity, because in this configuration no resonances can be excited in the antenna elements in the probed spectral range. Note that the overall density of plasmonic structures in the antenna arrays is small, resulting in relatively shallow resonances. In Fig.~\ref{characterisation}(b) we plot the corresponding data for the reference structures without tapers fabricated with the same exposure dose (sample v). Again we find a resonance for the incident light polarized in $x$-direction, which however, has a qualitatively different shape: it is narrower and more pronounced as compared to the tapered case. For $y$-polarization the transmittance is again close to unity. In order to check if the nanoantenna's functionality is sensitive on the width of the nanorods, we measured the transmittance spectra for both tapered and untapered antenna arrays with different nanorod diameters. These results are shown in Figs.~\ref{characterisation}(c,d) and show a clear red shift of the resonance position with decreasing the nanorod diameter both for the tapered nanoantennas and for the reference structures. Despite this resonance shift the observed resonance line-shapes and the strengths are preserved.

\section{Numerical simulations}
\vspace{-1mm}
To understand better the observed spectral features of the nanoantennas next we perform full-vectorial 3D simulations for our experimental parameters using CST Microwave Studio - finite integration technique. The nanoantennas were modeled as gold structures sitting on an ITO coated glass substrate of infinite depth. We use periodic boundary conditions applied to a unit cell composed of a single nanoantenna. In all simulations, plane waves with the electric field orientated either along \textit{x}- or along \textit{y}-direction were normally incident on the substrate and transmission coefficients describing the amplitude of the transmitted waves were calculated. The frequency-dependent refractive index of gold was taken from the literature~\cite{palik}. The refractive indices of the ITO and glass were chosen as $1.8$ and $1.5$, respectively, and assumed to be constant within the entire spectral range.

Due to shape deviations of the experimental nanorods from the ideal block shape and other sample imperfections the measured width of the fabricated nanorods may differ from those chosen in simulations. In order to facilitate \emph{qualitative} analysis, we performed a calibration simulation correlating experimental and theoretical transmittance spectra of a test grating consisting of a large number of equally spaced single gold nanorods. In these simulations we have varied the width of the nanorods in a wide range around the nominal value while leaving all other parameters unchanged ($L_f=235\,\rm nm$, $w+a=80\rm\,nm$, $h=50\,\rm nm$). We found that a good agreement between the experimental and theoretical values of the resonance frequency for sample vi can be achieved for $a\approx10\,\rm nm$.

\begin{figure}[t]
  \centering
  \includegraphics[width=\textwidth]{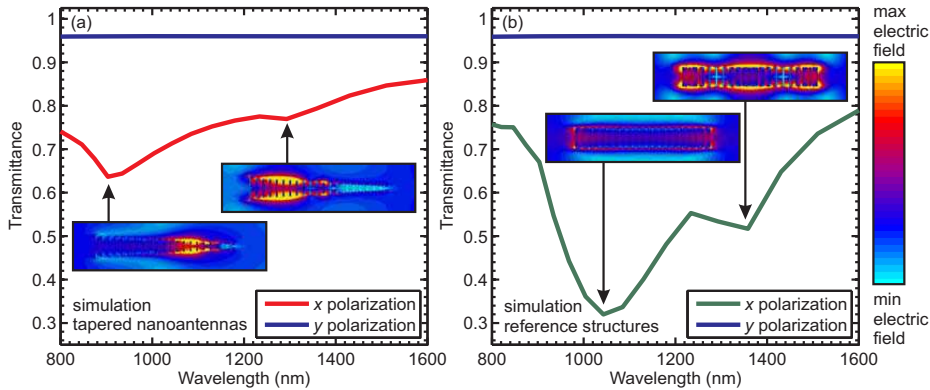}
  \vspace{-2mm}
  \caption{Calculated transmittance spectra of (a) tapered nanoantennas and (b) reference structures for $x$- and $y$-polarized light. The insets show the magnitude of the electric field vector in the $z$-plane of the nanoantenna at the spectral positions indicated by the black arrows.}
  \vspace{-3mm}
  \label{theory}
\end{figure}

Figure~\ref{theory} shows the simulated transmittance spectra of the tapered nanoantennas (a) and reference structures (b). In accord with our theoretical findings for the test grating, the width of the nanorods used in the simulations was set to 10\,nm. For \textit{x}-polarization we observe that the positions and the ratio of the depths of the resonances are in good agreement with those observed in the experiment [Fig.~\ref{characterisation}(a,b)]. Good qualitative agreement is also found between the theoretical and the experimental spectra for \textit{y}-polarization.

First we discuss the origin of the resonances for \textit{x}-polarization. In good agreement with the experiments [Fig.~\ref{characterisation}(b)], in Fig.~\ref{theory}(b) we observe that the reference structure is resonant around the single wavelength of $1040$\,nm. A similar resonant behavior was predicted for arrayed nanoantennas composed of identical plasmonic nanoparticles (see e.g., Ref.~\cite{pel_09}). A qualitatively different behavior is, on the other hand, observed for the tapered nanoantenna. Nanorods of different length resonate at different wavelengths and the overlap of their spectra results in a broad and shallow transmittance curve [Fig.~\ref{theory}(a)].

For both structures a double-dip in the transmittance line-shape is observed in the calculated spectra for $x$-polarization. A similar behavior was recognized in Ref.~\cite{maksymov_kivshar_2012b}, where individual nanorods of a tapered nanoantenna were excited simultaneously in phase by dipole emitters from the near-field zone. To unveil the origin of this double-dip shape of the transmittance curves we calculated the magnitude of the electric field vector in the $z$-plane at the center of the gold nanorods for those excitation wavelengths for which the minima are observed. These results are shown in the insets of Fig.~\ref{theory}. Obviously, the double-dip line shape originates through completely different mechanisms in the two structures. For the tapered nanoantenna the electric fields are tightly concentrated at opposite ends of the structure for the two different minima. While it is evident that individual nanoantennas arranged in a matrix might act as an effective plasmonic grating, this result clearly indicates that in the present case the transmittance line shape is not dominated by collective effects in the spectral range of interest but can be attributed to different modes in isolated nanoantennas. We believe that the interaction between the individual tapered nanoantennas in \textit{x}-direction may be neglected since the nanoantennas are designed to emit in \textit{y}-direction only, and because the electromagnetic field is tightly confined to the edges of their nanorods~\cite{mak11_1, mir11, maksymov_kivshar_2012b}. Most importantly, this double-dip spectrum, which is also visible in the experimental data for the tapered nanoantennas [compare Fig.~\ref{characterisation}(c)] is directly connected to the nanoantennas' broadband functionality and the decrease of the nanoantenna's effective length with decreasing wavelength, as discussed in more detail in Ref.~\cite{mir11}.

For the reference structure, on the other hand, no such wavelength dependent spatial light concentration can be identified. Instead, the electric-field profiles suggest that the double-dip line shape is due to a Fabry-Perot type resonance caused by reflections on the edges of the finite plasmonic chain. This explanation is backed up by the fact that we do not observe a double dip line shape in our simulations for an infinite plasmonic chain. In the experimental spectra only a slight distortion of the transmittance line shape from a typical resonance shape is found, as best visible in the blue curve of Fig.~\ref{characterisation}(d).

Our findings are similar to the previous theoretical results for single tapered nanoantennas without a substrate~\cite{mir11, maksymov_kivshar_2012b}. Importantly, the presence of the substrate does not affect the broadband optical properties of tapered nanoantennas. It merely deviates their directional emission toward the optically denser substrate~\cite{curto_vanhulst_2010} and causes a red shift in operation frequency.

\section{Conclusions}
\vspace{-1mm}

We have studied experimentally a novel type of optical nanoantennas with strong directionality and large spectral bandwidth based on tapered Yagi-Uda nanoantennas. We have fabricated and optically characterized arrays of tapered optical nanoantennas. We have observed that measured transmittance spectra reflect the broadband characteristics of the tapered nanoantennas, which originates from the overlap of the resonances of the individual nanorods, and reveal a fundamentally different behavior as compared to their untapered counterparts, arrays of identical nanorods. Our results are in a good agreement with the full-vectorial numerical calculations which further demonstrate the origin of the observed spectral features.

\section{Acknowledgments}
\vspace{-1mm}
We acknowledge a support from the Australian National Fabrication Facility and the Australian Research Council through Future Fellowship, Discovery, and Centre of Excellence projects.

\end{document}